\documentclass[12pt]{article}

\usepackage{a41}
\usepackage{floatfig,epsfig}
\usepackage{rotating}

\newcounter{my}

\newcommand{\la}[1]{\label{#1}}
\newcommand{\re}[1]{\ (\ref{#1})}
\newcommand{\nn}{\nonumber}
\newcommand{\ed}{\end{document}}
\newcommand{\be}{\begin{equation}}
\newcommand{\ee}{\end{equation}}

\newcommand{\ba}{\begin{eqnarray}}
\newcommand{\ea}{\end{eqnarray}}
\newcommand{\baz}{\begin{eqnarray*}}
\newcommand{\eaz}{\end{eqnarray*}}
\newcommand{\bb}{\begin{thebibliography}}
\newcommand{\eb}{\end{thebibliography}}
\newcommand{\ct}[1]{${\cite{#1}}$}

\newcommand{\bi}[1]{\bibitem{#1}}

\begin{document}
\initfloatingfigs
\sloppy
\thispagestyle{empty}

\vspace{1cm}

\mbox{}

\vspace{2cm}

\begin{center}
{\Large {\bf  Instantons,  Spin Crisis \\
 and   \\ [2mm]
High $Q^2$ Anomaly at HERA}}
\footnote{ The talk presented at the
Workshop "Physics with Polarized Protons at HERA",
August 1997, DESY-Zeuthen}\\[2cm]
{\large N.I.Kochelev}\\[0.2cm]
{\it
 Bogoliubov Laboratory of Theoretical Physics,\\
Joint Institute for Nuclear Research,\\
RU-141980 Dubna, Moscow region, Russia}\\ [0.3cm]

\end{center}
\vspace{2cm}

\begin{abstract}
\noindent
The contribution of the nonperturbative quark-gluon   interaction
induced by the instantons  to   $g_1^p(x,Q^2)$ and
$F_2^p(x,Q^2)$ structure functions is estimated.
It is shown that nontrivial $Q^2$ dependence of the instanton
contribution to $g_1^p(x,Q^2)$ allows us to explain the observed
decreasing of the part of the proton spin carried
by quarks  without involving a large
positive gluon polarization.

It is demonstrated that the anomalous enhancement of the instanton
contribution to $F_2^p(x,Q^2)$ structure function, due to multiple
emission of the gluons from instanton vertex, can be one of the reasons
 for the excess of the DIS events at HERA
at high $Q^2 $ and $x$.
\end{abstract}

\newpage
\section{Introduction}
\vspace{1mm}
\noindent
At the present time the problem of the proton spin is one of the most
exciting problem in QCD.
This problem was inspired by the result of the EMC \ct{EMC}, which found
out that only  the small amount of proton spin is carried by its quarks.
This result has been confirmed by more accurate experiments
\ct{Conf}.

In spite of many interesting ideas,  which have been proposed
to explain  this surprising result,
 the mechanism  responsible for the decreasing of the
value of helicity carried by quarks, is still not clear
(see recent review \ct{AA}).

A new  challenge for Standard Model  arose from the recent HERA
result for unpolarized DIS events at high $x$ and $Q^2$ \ct{HERA}.
Two collaborations, H1 and ZEUS, observed the large excess of high $Q^2$
and $x$
DIS events in contradiction with the Standard Model expectations.
  Many interesting explanations of the HERA anomaly
beyond Standard Model have been suggested (see for example \ct{leptq}).

However, from our point of view, in the  analysis of
these experimental data, the  very important feature of  QCD
has not been taken into account. This particularity is connected
with the possibility of new type of the nonperturbative quark-gluon
interaction \ct{koch1}, which can be
induced by strong vacuum fluctuations of
the gluon fields,
so-called instantons \ct{Pol}. These instanton fluctuations
describe the quantum tunneling between different gauge rotated classical
vacua in QCD and give  us the very important information on the
complicated structure of the ground state in theory of strong interaction.

In this article we   estimate  the contribution from the instanton
induced quark-gluon chromomagnetic interaction \ct{koch1}  to the
$g_1^p(x)$ and $F_2^p(x)$ structure functions
 in instanton liquid model for QCD vacuum \ct{a5}, \ct{a55}.

We show that this  contribution
to quark depolarization allows to explain the observed decreasing
 of the spin-dependent structure function
$g_1^p(x)$ at $x>0.001$.

The possibility to explain the excess of high $Q^2$ and large $x$ events
at HERA
by
the  anomalous enhancement of the instanton contribution to
 structure function $F_2^p(x,Q^2)$ is shown.

\section{ Anomalous Quark-Gluon Interaction
 Induced by Instantons}
\vspace{1mm}
\noindent
One of the models for the description of  nonperturbative
effects  in QCD is the instanton liquid model (see reviews \ct{a5},
\ct{a55}). In the framework of this model many fundamental quantities of
the QCD vacuum,  such as  different types of the quark and gluon
condensates were described rather well.
Furthermore,  this model also  gives  good description of the
important hadron properties, e.g. masses,
decay widths, form factors  etc.

The existence of  instantons leads to a  specific
 quark-quark  and quark-gluon interaction through the QCD vacuum,
which
has the following form \ct{CDG}

\ba
{\cal L}_{eff}&=&\int\prod_q(m_q\rho-2\pi^2\rho^3\bar q_R(1+\frac{i}{4}
\tau^aU_{aa^\prime}\bar\eta_{a^\prime\mu\nu}\sigma_{\mu\nu})q_L)
\nn\\
&\cdot &exp^{-\frac{2\pi^2}{g}\rho^2U_{bb^{\prime}}
\bar\eta_{b^\prime\gamma\delta}
G^b_{\gamma\delta}}
\frac{d\rho}{\rho^5}d_0(\rho)d\hat{o}
+R\longleftrightarrow L,
\label{e1}
\ea
 where $\rho$ is the instanton size, $\tau^a$ are the matrices of the
 $SU(2)_c$ subgroup of the $SU(3)_c$ colour group,
 $d_0(\rho)$ is the density of the instantons, $d\hat{o}$ stands
 for integration over the instanton orientation in colour space,
$\int d\hat{o}=1$,
$U$ is the orientation matrix of the instanton,
 $\bar\eta_{a\mu\nu}$ is the numerical t'Hooft symbol and
 $\sigma_{\mu\nu}=[\gamma_\mu,\gamma_\nu]/2$.

From this Lagrangian one can find the famous t'Hooft
{\it quark-quark} interaction \ct{Hooft}, which is a corner stone for
many applications of the instanton physics.

Recently, it was shown that from Eq.\re{e1}
 a new type of the nonperturbative  quark-gluon  interaction can be
 obtained. This interaction has the form of
 {\it anomalous chromomagnetic
quark-gluon} interaction \ct{koch1}
\be
\Delta {\cal L_A}=
-i\mu_a
\sum_q\frac{g}{2m_q^*}\bar q\sigma_{\mu\nu}
t^a qG_{\mu\nu}^a.
\label{e4}
\ee
The value of the quark anomalous chromomagnetic moment
in the liquid instanton model is
\be
\mu_a=-\frac{f\pi}{2\alpha_s},
\label{a6}
\ee
 where $f=n_c\pi^2\rho_c^4$ is the so-called packing fraction of instantons
 in  vacuum,
\\$m_q^*=m_q-2\pi^2\rho_c^2<0\mid \bar qq\mid 0>/3$ is
the effective quark mass.
The value of $n_c$ is connected with the value of the
gluon condensate by the formula:
\be
n_c=<0\mid \alpha_sG_{\mu\nu}^a G_{\mu\nu}^a\mid 0>/16\pi
\approx 7.5~10^{-4}{\  }GeV^4.
\nn
\ee
The following estimate for
 the value of the anomalous quark chromomagnetic
moment has been obtained for $\rho_c=1.6GeV^{-1}$ in \ct{koch1}
\begin{equation}
\mu_a=-0.2.
\nonumber
\end{equation}

The principal difference between the instanton induced interactions
\re{e1},\re{e4} and the perturbative quark-gluon vertex
\be
{\cal L}_{pert}=g\bar q\gamma_\mu t^aqA_\mu^a,
\la{pert}
\ee
is the  quark helicity flip at instanton vertex.
Therefore this interaction  can give a contribution to different
spin-dependent cross sections, in particular to the spin-dependent
structure function $g_1(x,Q^2)$.

\section{Instantons  and "Spin Crisis"}
\vspace{1mm}
\noindent
The diagram, which gives rise to
   the  {\it quark} structure
functions from interaction \re{e4}
is presented in  Fig.1a.
\begin{figure}[htb]
\centering
\epsfig{file=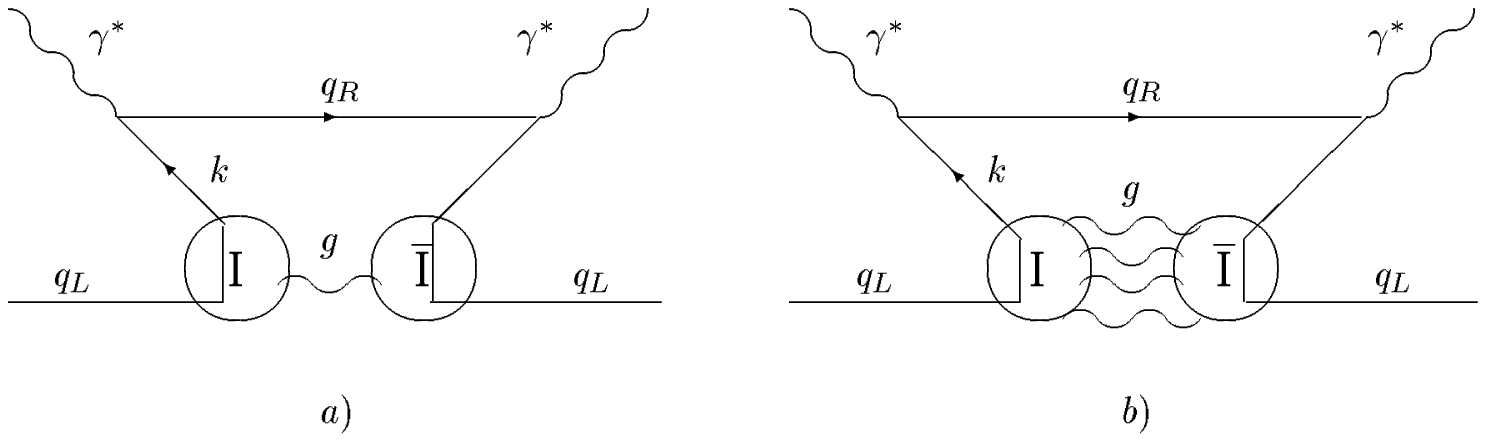,width=9cm}
\vskip 3cm
\caption{\it The instanton contributions to  proton
structure function from the anomalous
chromomagnetic interaction (a), and  from the multiple
gluons emission from instanton vertex (b). The label $I(\bar I)$ denotes
instanton(antiinstanton).}
\end{figure}
 The contribution
to quark spin-dependent structure function $g^q_1(x,Q^2)$ can be
obtained by the
projection of the imaginary part of the forward Compton scattering
amplitude $T_{\mu\nu}$
\be
g^q_1(x,Q^2)=-\frac{ie_{\mu\nu\rho\sigma}p^\rho q^\sigma ImT_{\mu\nu}}{2p.q},
\la{pro}
\ee
where $p$ is the momentum  of the initial valence
quark in nucleon.  The straightforward  calculation of the contribution of
diagram Fig.1a leads to the result
\be
g_1^q(x,Q^2)=-\frac{e_q^2}{8}\frac{|\mu_a|\rho_c^2}{(1-x)}
\int_0^{\frac{Q^2(1-x)}{4x}}dk_\bot^2\\ \nn
\frac{F^2(k\rho_c/2)}{\sqrt{1-\frac{4xk_\bot^2}{(1-x)Q^2}}},
\la{fig1a}
\ee
where
\be
F(z)=z\frac{d}{dz}[I_0(z)K_0(z)-I_1(z)K_1(z)],
\la{ff}
\ee
is the Fourier transformation of the quark zero mode in instanton field,
 $k^2=k_\bot^2/(1-x)$,
 $x=Q^2/2p.q$
and the relation
$\alpha_s\mu_a^2/m_q^{*2}\rho_c^2=3\pi|\mu_a|/8$ \ct{pol}
has been used.

The  very interesting feature of the instanton contribution
\re{fig1a} is its specific $Q^2$ dependence.
Namely, at small $Q^2\ll 1/\rho_c^2$ it is proportional to $Q^2$  and for
$Q^2\gg 1/\rho_c^2$ it is the constant.
 Therefore, the $Q^2$ dependence of the instanton
contribution to polarized   structure functions
{\it should be different} from the ordinary perturbative $Log(Q^2/\Lambda^2)$
evolution.  Due to  definite value of the quark helicity induced by the
instantons, the same conclusion applies to the instanton
 contribution
to the unpolarized structure functions as well. The fundamental
reason for such a behavior
 is the quark spin-flip induced by instanton,
which adds the extra power of $k_\bot$ to the matrix element for forward
Compton scattering amplitude.
 As result, at $Q^2=0$ the instanton
contribution to $g_1(x,Q^2)$ is {\it zero}. Therefore the value of the
helicity  of quarks  at $Q^2\gg 1/\rho_c^2$ {\it should
 differ} from the  value of helicity carried by  proton quarks, that was
extracted from baryon spectroscopy at very small $Q^2$.
This  anomalous $Q^2$  dependence of the instanton contribution to
quark polarization  can be one of the
reasons for the success of the simple three quark constituent model in
the description of  nucleon magnetic moments at $Q^2=0$.

In the large $Q^2$ limit the contribution to $g_1^q(x)$ is constant
\be
g_1^q(x)=-\frac{e_q^2}{4}|\mu_a|.
\la{ff1}
\ee
It should be mentioned that the sign of the correction to $g_1^q$ structure
function is {\it negative} and comes from
{\it negative} valence quark polarization induced by instantons
inside proton (Fig.1).  Furthermore, this contribution does not
depend on $x$ and therefore can give rise to proton
structure functions at rather large values of Bjorken variable $x$.

To estimate the contribution to the proton structure functions,
we will use the simple convolution model
\be
g_1^p(x)=\sum_q\int_x^1\frac{dy}{y}g_1^q(\frac{x}{y})\Delta q_V(y),
\nn
\ee
where $\Delta q_V(y)$ are the initial valence quark polarizations,
which are taken in the form
 \ba
\Delta u_V(x)&=&3.7(1-x)^3, {\ }\Delta d_V(x)=-1.3(1-x)^3,\label{val}
\ea
 normalized
 to the experimental data
on the weak decay coupling constants of  hyperons
\be
g_A^3=\Delta u_V-\Delta d_V=1.25; {\ } g_A^8=\Delta u_V+\Delta d_V=0.6.
\label{coupl}
\ee

In Fig.2 the result of the  calculation of instanton contribution to
$g_1^p(x)$ in the region of Bjorken variable  $0.001< x<1$
is presented.
\begin{figure}[htb]
%\begin{figure}
\centering
\epsfig{file=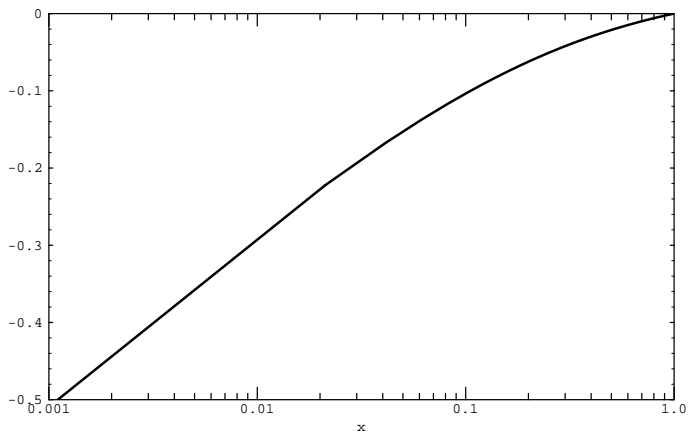,width=9cm}
\vskip -3cm
\caption{\it  The ratio of the contributions from instanton induced
quark depolarization and
initial valence quark polarization
 to proton spin-dependent structure function
$g_1^p(x)$. }
\end{figure}
As we see, the quark depolarization induced by the quark-gluon
interaction through instantons  leads
to a rather strong decrease
of the value of the $g_1^p(x)$, in particularly at large
values of $x$.

Therefore, the considerable  part of the observed
decreasing of the part of the proton spin carried
by quarks  can be explained by the  contribution from anomalous
quark-gluon chromomagnetic interaction induced by instantons.
The additional decreasing of the value of the proton helicity
carried by quarks, especially at low $x$ region, can be connected
with the contribution to quark
depolarization which comes from
the quark-quark  t'Hooft \ct{Hooft} interaction induced by instantons
\ct{koch} (see also \ct{DorKoch}, \ct{Forte}).

The instanton mechanism of the "spin crisis" solution is opposite
to the approach  which is based on the  assumption about
the large {\it positive} gluon polarization inside proton \ct{AA}.
It was shown recently \ct{pol} that instantons lead to
{\it a negative} gluon polarization and therefore instanton model
for QCD vacuum rules out the mechanism based on
positive gluon polarization.

\section{Instantons and High $Q^2$ anomaly at HERA}
\vspace{1mm}
\noindent
The contribution of instantons to the {\it unpolarized}  structure function
$F_2^P(x,Q^2)$
can be estimated in the similar way. Due to definite helicity
of the quarks in instanton field (see Fig.1a), the instanton
contributions to quark structure functions $f_1^q(x)$ and $g_1^q(x)$
are related to each other by the following expression
\be
f_1^{q}(x,Q^2)=-g_1^{q}(x,Q^2),
\nn
\ee
The proton structure function is given
by the formula
\be
F_1^{P,Inst}(x,Q^2)=\int_x^1\frac{dy}{y}f_1^{q}
(\frac{x}{y},Q^2)q_V(y,Q^2),
\nn
\ee
where valence unpolarized distribution function $q_V$
is taken in the simple form
\be
q_V(x)=Ax^{-0.5}(1-x)^3
\nn
\ee
with constant $A$ normalized to the number of $u-$ and $d-$ quarks in proton.

The result of the  calculation of the instanton contribution
 to $F_2^p(x)$
in the region  $0.1< x<1$
is shown in Fig.3.
\begin{figure}[htb]
%\begin{figure}
\centering
\epsfig{file=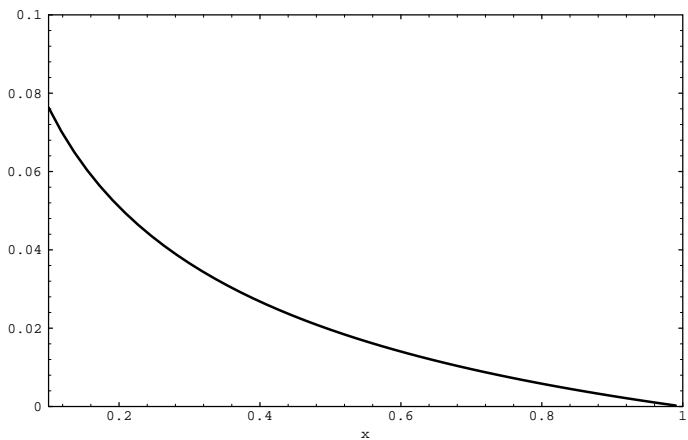,width=9cm}
\vskip -3cm
\caption{\it  The ratio of contributions from the instantons  and
  valence quarks to proton  structure function
$F_2^p(x)$. }
\end{figure}
This contribution is rather large $\approx 5\%$ in this region and
therefore should
%\newpage
 be taken
into account in the analysis of $Q^2$ dependence of $F_2(x,Q^2)$
\footnote{ In \ct{br},\ct{ring}
some contribution from
the instantons to the coefficient functions of the structure
function of unpolarized DIS, which was connected
with exponentially suppressed on $Q^2$ part of the quark propagator
in instanton field, was calculated
in the dilute instanton gas approximation.
The very small value, which was found in \ct{br},\ct{ring},
comes from  the  cut-off of the contribution from
the instantons with the large size $\rho\ge1/Q$.}.
 The special
 interest is the different $Q^2$ dependence of the instanton and
perturbative gluon contributions to $F_2(x,Q^2)$,
which was mentioned above.

The first results of search for the instanton induced events
in DIS  have been published by  H1 Collaboration \ct{H1}.
This collaboration  obtained very small upper limit for the instanton
contribution to structure function $F_2^p(x,Q^2)$.
However, this limit was obtained without taking into account
the possible features of the hadronization of the soft gluons from instanton
vertex. One of such features is the strong  Mueller's
interaction of the  gluons \ct{mull} and therefore
the assumption on the independent fragmentation of the gluons
emitted from instanton, which was used in \ct{H1},
is probably incorrect.

Recently, the very interesting results on the structure function
 $F_2^p(x,Q^2)$  at high values
of $x$ and $Q^2$ from HERA have been published \ct{HERA}. The analysis of
 the data,
which was based on {\it the assumption} of its perturbative
$Log(Q^2/\Lambda^2)$ evolution of quark distribution functions
 in this
region, discovered  large excess of events, as compared with perturbative
QCD predictions.

From our point of view, the natural explanation of this excess
can be the anomalous enhancement  of the instanton contribution
to $F_2^p(x,Q^2)$ structure function at high $Q^2$ and large $x$
region, which comes from multiple creation of soft gluons from
instanton vertex Fig.1b.

It should be mentioned that the similar mechanism of enhancement of
the electro-weak instantons
 contribution  to the cross  sections with violation of the baryon
number conservation
due to multiple creation of gauge bosons by instanton,
is widely discussed \ct{ring1}.

We can estimate the factor of enhancement in the case of the large
 number
 of the created gluons.
When the distance between $I$ and $\bar I$ is large,
$R_{I\bar I}\gg\rho_c$,  and
  $n_g\to\infty$, the correction
can be written  by formula \ct{br}
\be
F_{enh}\approx exp({-\delta S_{I\bar I}})\approx
exp(\frac{24\pi}{\alpha_s(\rho_c)}\frac{1}{\xi^2}),
\la{enc}
\ee
where
$\xi$ is the so-called  conformal parameter
\be
\xi=\frac{R_{I\bar I}^2+2\rho_c^2}{\rho_c^2},
\nn
\ee
and $\delta S_{I\bar I}$  is the variation of the value of the action
for the instanton-antiinstanton configuration due to the dipole-dipole
$I\bar I$ interaction.
For  $\alpha_s(\rho_c)\approx 0.36$ \ct{koch1}, and
$R_{I\bar I}\approx 3\rho_c $
 \ct{a5}, the enhancement is
$F_{enh}\approx 10!!!$.  Therefore we should have approximately
 $50\%$ larger value of $F_2^p(x,Q^2)$ at $x>0.1 $ as compared with
the perturbative QCD predictions if the large number of the
  gluons can be created.

This enhancement  comes from the large phase space
allowed for gluons, which are created by the instanton in $\gamma^*(Z,W)q$
collision at high energy. So, one can write for Bjorken variable
$x$ the following formula $x=Q^2/(M_X^2+Q^2)$, where
 $M_X$ is the mass of the produced
hadron system.
The value of  energy of the gluon that is emitted by instanton
 equals approximately to \\$E_g\approx 1/\rho_c\approx 1GeV$.
 At some fixed value of $x$, for example at $x\approx 0.5$, and
high $Q^2\approx 10^4 GeV^2$, where excess of events at HERA was found,
 we have $M_X\approx 100 GeV $ and therefore
 a large amount of
  gluons can be emitted. In this case the large enhancement of
 instanton contribution takes place. On the other hand, at
$Q^2\approx 10 GeV^2$, and
high $x>0.5$ region , where the good experimental data on $F_2(x,Q^2)$ are
available,  $M_X<10 GeV$ and  as the result
the number of the possible additional gluons is rather small. Therefore at
low $Q^2$ and large $x$ the additional enhancement  of the instanton
contribution to structure functions is absent.

The instanton  mechanism for the high $Q^2$ anomaly at HERA can be checked
 by using the  future possible
polarized option at HERA, because strong helicity dependence
of the instanton contribution should lead to  large double spin
asymmetry of these events. For example, due to negative quark polarization
induced by instantons, at high $x$ and $Q^2$ region  we expect
  large {\it negative } double spin asymmetry $A_L$ for the cross sections in
 the scattering of
longitudinal polarized lepton and hadron
beams at HERA. This prediction is opposite to the
positive value of the asymmetry, which could be expected
if  the perturbative QCD works in this kinematical region.

\section{Summary}
\vspace{1mm}
\noindent
In summary,  the instanton induced  quark-gluon interaction
leads  to a {\it large negative} contribution to the proton
spin-dependent structure
function $g_1^p(x,Q^2)$. This allows us to explain the observed decreasing
of the $g_1^p(x,Q^2)$ structure function in comparison with
the  prediction of the naive quark model.

It is shown that the same interaction can be responsible for the
high $Q^2$ excess of DIS events at HERA.

Thus, from our point of view,  these two phenomena, ''spin crisis'' and
 high $Q^2$ and $x$  anomaly at HERA, can be the first direct
manifestation of the complicated
structure of QCD vacuum in deep inelastic scattering:  a vacuum filled with
intensive nonperturbative fluctuations of gluon fields, instantons.

The future polarized beams at
HERA can be a unique place for investigation of the instanton induced
events in deep-inelastic scattering because it will allow us to investigate
the strong spin and $Q^2$ dependence of the instanton contribution
to structure functions in  the very  wide kinematic region.

\section{Acknowledgements}
\vspace{1mm}
\noindent
The author is
thankful to J.Bl\"umlein, S.Brodsky, J.Dainton, A.E.Dorokhov, A.V.Efremov,
E.Elsen, M.Erdmann, F.Jegerlehner, T.Morii, P.S\"oding
and all participants of
the Workshop ``Physics with Polarized Protons at HERA'' and
especially to  A. De Roeck
for many stimulated discussions.

This work was supported in part by the Heisenberg-Landau program and by the
Russian Foundation for Fundamental Research (RFFR) 96-02-18096.

\end{document}